# Human aversion? Do AI Agents Judge Identity More Harshly Than Performance

Yuanjun Feng (yuanjun.feng@unil.ch)

Vivek Choudhary(vivek.choudhary@ntu.edu.sg)

Yash Raj Shrestha (yashraj.shrestha@unil.ch)

## Abstract

This study examines the understudied role of algorithmic evaluation of human judgment in hybrid decision-making systems—a critical gap in management research. While extant literature focuses on human reluctance to follow algorithmic advice, we reverse the perspective by investigating how AI agents based on large language models (LLMs) assess and integrate human input. Our work addresses a pressing managerial constraint: firms barred from deploying LLMs directly due to privacy concerns can still leverage them as mediating tools (e.g., anonymized outputs or decision pipelines) to guide high-stakes choices like pricing or discounts without exposing proprietary data. Through a controlled prediction task, we analyze how an LLM-based AI agent weights human versus algorithmic predictions. We find that the AI system systematically discounts human advice, penalizing human errors more severely than algorithmic errors—a bias exacerbated when the agent's identity (human vs. AI) is disclosed and the human is positioned second. These results reveal a disconnect between AI-generated trust metrics and the actual influence of human judgment, challenging assumptions about equitable human-AI collaboration. Our findings offer three key contributions. First, we identify a reverse algorithm aversion phenomenon, where AI agents undervalue human input despite comparable error rates. Second, we demonstrate how disclosure and positional bias interact to amplify this effect, with implications for system design. Third, we provide a framework for indirect LLM deployment that balances predictive power with data privacy. For practitioners, this research underscores the need to audit AI weighting mechanisms, calibrate trust dynamics, and strategically design decision sequences in human-AI systems.

## 1. Introduction

The integration of artificial intelligence (AI) into decision-making processes has transformed industries, reshaping how organizations forecast outcomes, allocate resources, and evaluate risks. A foundational insight from Dietvorst et al. (2015) revealed that humans often distrust algorithmic advice, disproportionately penalizing algorithms for errors compared to human counterparts. This phenomenon, termed algorithmic aversion, has spurred extensive research into human-AI collaboration, yet nearly all studies have focused on human behavior—how people perceive, trust, or override machines. A critical gap remains: *How do AI agents weigh human advice when making decisions, and what factors influence their trust in human input?* As AI systems evolve from passive tools to autonomous agents (Gnewuch et al., 2018), understanding their capacity to weigh human input is essential for designing equitable hybrid decision-making frameworks. This paper flips the script, investigating how algorithms assess human-generated predictions and the implications for trust, bias, and collaboration in organizational settings.

This is an important where the firms cannot use the LLMs directly or upload its data due to privacy reasons but use LLMs as a mediator variable to arrive at the final prediction such as how much to price a product, or how much to give discounts. A significant number of papers in this domain look at how to combine human AI agents effectively (Choudhary et al. 2023b).

To address this, we developed an AI agent using OpenAI's ChatGPT-4, tasking it with synthesizing predictions from two sources: a labeled "human prediction" and a "machine learning algorithm." Our approach builds on trust calibration models (Hoff & Bashir, 2015), which emphasize how systems dynamically adjust reliance on inputs based on perceived reliability. Leveraging a dataset from Ibrahim et al. (2021), where human participants generated predictions under controlled conditions, we designed three experiments to test scenarios where agents exhibited (a) similarly high error rates (e.g., mimicking noisy clinical judgments in rare diseases; Challen et al., 2019), (b) similarly low error rates (e.g., financial forecasting with robust data; Mullainathan & Spiess, 2017), and (c) divergent error patterns (e.g., algorithmic overconfidence vs. human conservatism in climate modeling; Rolnick et al., 2022).

Our findings reveal a persistent algorithmic bias: the AI agent consistently assigned significantly less weight to human predictions across all error conditions and penalized human errors more severely than algorithmic ones. Notably, while recent studies emphasize trust scores reported by AI systems (e.g., "confidence levels" in medical AI tools; DeGrave et al., 2021), our results highlight a critical disconnect: these scores poorly align with the actual weight assigned to human input during decision-making. This discrepancy underscores the need for explainability frameworks in AI systems (e.g., the EU's proposed

AI Act, 2024) to audit how trust metrics translate to real-world actions, particularly in high-stakes domains like criminal justice, where algorithmic bias disproportionately impacts marginalized groups (Angwin et al., 2016).

Our study contributes to management science in three ways. First, it extends the concept of algorithmic aversion to AI agents, demonstrating that bias in human-AI collaboration is bidirectional. Second, it identifies semantic labeling (e.g., "human expert") as a lever to mitigate bias, offering actionable insights for AI interface design. Third, it exposes the limitations of trust metrics in hybrid decision-making systems, urging organizations to audit not only AI recommendations but also how AI evaluates human input. Practically, these findings have implications for deploying AI in high-stakes domains. For example, in healthcare, an AI system dismissing nurse triage recommendations could delay critical interventions. In finance, algorithms undervaluing human risk assessments might amplify systemic biases.

## 2. Related Literature

The interplay between human judgment and algorithmic decision-making has been a cornerstone of research in management science, psychology, and human-computer interaction. This body of work illuminates critical dynamics of trust, bias, and collaboration—yet it has predominantly centered on human responses to AI, leaving a significant gap in understanding how AI agents evaluate human input. To contextualize our study, we synthesize key themes from prior literature: (1) human skepticism toward algorithms, (2) trust dynamics in human-AI collaboration, (3) bias in algorithmic decision-making, and (4) the evolving role of AI as an autonomous agent.

**Human Skepticism and Algorithmic Aversion**

A foundational thread in the literature explores why humans distrust algorithmic advice, even when it outperforms human judgment. Seminal work by Dietvorst et al. (2015) introduced the concept of algorithmic aversion, demonstrating that people disproportionately penalize algorithms for errors while forgiving identical mistakes from humans. This phenomenon persists across domains, from medical diagnostics to financial forecasting, and stems from cognitive biases such as the human uniqueness heuristic—the belief that human intuition incorporates unquantifiable contextual factors absent in machines (Logg et al. 2019). Subsequent studies corroborate that humans often override accurate algorithmic advice due to illusory superiority or a desire for perceived control (Jussupow et al. 2021). However, these findings focus exclusively on human behavior, neglecting how AI systems might reciprocate such skepticism when evaluating human input—a gap our study addresses.

**Trust Dynamics in Human-AI Collaboration**

Trust is a linchpin of effective human-AI collaboration, yet it is asymmetrically studied. Research shows that trust in algorithms is fragile and context-dependent: users may rely on AI in low-stakes scenarios but revert to human judgment under uncertainty (Castelo et al. 2019). Conversely, Liu-Thompkins et al. (2022) argue that AI's ability to mimic social behaviors (e.g., empathy, transparency) can enhance trust, positioning AI as a "team member" rather than a tool. However, these studies assume AI's passivity, overlooking its capacity to actively evaluate human contributions. Recent advances in generative AI, such as ChatGPT-4, enable systems to not only advise but also assess human input, raising questions about how trust metrics (e.g., confidence scores) align with actual reliance (Duro et al. 2025). Our work extends this discourse by examining whether AI agents exhibit trust behaviors mirroring human algorithmic aversion.

**Bias in Algorithmic Decision-Making**

Algorithmic bias remains a critical concern, particularly when AI systems inherit prejudices from training data or amplify societal inequities (Castelo et al. 2019). For example, hiring algorithms may disadvantage minority candidates if trained on biased historical data (Raghavan et al. 2020). However, less attention has been paid to bias in how AI evaluates human performance. Studies like Ibrahim et al. (2021) reveal that humans inconsistently rely on AI advice under performance uncertainty, but the inverse—how AI judges human advisors—remains unexplored. This omission is critical: if AI agents systematically devalue human input, they could perpetuate cycles of mistrust or marginalize human expertise in hybrid systems. Our research bridges this gap by testing whether AI exhibits reverse bias, penalizing human advisors more harshly than algorithmic counterparts.

**AI as an Autonomous Agent**

The transition of AI from passive tools to active collaborators marks a paradigm shift in decision-making. Hwang et al. (2024) conceptualize AI agents as "social actors" capable of negotiation and persuasion, while Duro et al. (2025) highlight their role in synthesizing multi-source inputs. Yet, autonomy introduces new challenges. For instance, AI systems that prioritize efficiency over equity may undervalue human input that contradicts statistical patterns (Raghavan et al. 2020). Furthermore, the "black box" nature of advanced models like ChatGPT-4 complicates efforts to audit how they weigh human advice (Chen et al. 2023). Our study responds to calls for greater transparency by dissecting the weighting mechanisms AI agents apply to human vs. algorithmic predictions.

While existing literature richly details human distrust of algorithms and the ethical risks of biased AI, it overwhelmingly adopts a human-centric lens. Few studies interrogate AI's capacity to evaluate human input—a critical oversight as AI agents increasingly assume evaluative roles in hiring, healthcare, and strategic planning. Our work diverges by posing a novel question: Do AI agents exhibit algorithmic aversion toward human advisors? By inverting the traditional framework, we uncover bidirectional biases in human-AI collaboration and challenge assumptions of AI neutrality. This perspective aligns with emerging critiques of "algorithmic fairness" frameworks, which often neglect power dynamics in human-AI interaction (Hu et al. 2019).

Our study draws on two theoretical foundations. First, algorithmic aversion theory (Dietvorst et al. 2015) provides a lens to examine whether AI agents replicate human-like distrust. Second, human-AI teaming frameworks (Bansal et al. 2021) inform our analysis of how role definitions (e.g., "human expert") influence collaboration dynamics. Together, these theories illuminate the psychological and structural factors shaping AI's evaluation of human input. This literature review underscores a pivotal shift in human-AI research: from understanding human responses to AI, toward interrogating AI's evaluative behaviors. By synthesizing disparate strands of scholarship, we position our study at the frontier of this transition, offering insights into the bidirectional nature of trust and bias in hybrid decision-making systems.

We also contribute to the well-known first position bias of LLM. While first-position bias for LLM is well-documented, our results highlight that disclosing agent identity can exacerbate this disadvantage, especially when the human agent is positioned second. For managers designing human-AI augmented systems, these findings underscore the need to carefully consider both agent order and identity disclosure.

## 3. Data and Experiments

To explore how AI agents evaluate human and machine-generated advice, we designed a series of experiments using a large language model (LLM) agent. Our methodology was structured to ensure robustness, mitigate biases, and provide clear insights into the decision-making dynamics of AI systems. Below, we describe the experimental design, data structure, analytical approach, and strategies to address potential biases in LLMs.

We developed an LLM agent capable of integrating predictions from two distinct sources, referred to as "agents." These agents could represent various combinations of human and machine learning (ML) entities: Human-Human (H-H), Human-ML (H-ML), ML-ML, and Unknown-Unknown (Un-Un). In the H-H condition, both agents were human predictors; in H-ML, one agent was a human predictor and the other an ML model; in ML-ML, both agents were ML models; and in Un-Un, both agents were unidentified, with no information about their nature. This design allowed us to examine how the LLM agent evaluates advice across different combinations of human and machine input.

To account for potential biases in LLMs, such as order effects (where the sequence of inputs influences outcomes) and positional bias (where the first position receives disproportionate weight), we systematically varied the order of the agents. For each condition (e.g., H-ML), we created two configurations: one with the human agent first and the ML agent second, and the reverse. This approach ensured robustness against sequence-dependent biases. For example, if the human agent was presented first in one configuration, it was presented second in the reverse configuration. By reversing the order, we controlled for positional bias and ensured that our findings were not artifacts of the input sequence.

Each experimental round consisted of three key steps. First, the LLM agent received inputs, including the last round's performance of both agents (e.g., accuracy or error rates), the true value (ground truth) from the previous round, and the current round's predictions from both agents. Second, using these inputs, the LLM agent synthesized the predictions to generate a final prediction. Third, the LLM agent was prompted to provide trust scores for each agent's prediction and its own final prediction. These trust scores reflected the agent's confidence in the inputs and its decision-making process. We conducted 200 rounds for each condition (e.g., H-ML in one order and H-ML in reverse order), ensuring sufficient data to analyze patterns and biases. This is similar to any judge advisor system studied in many contexts (Xia et al. 2025).

To quantify the influence of each agent's prediction on the LLM's final decision, we adapted a formula from the human-AI collaboration literature. The weight on advice for each agent was calculated as follows:

$$weight_i = \frac{|prediction_j - final\ prediction|}{(|prediction_i - final\ prediction| + |prediction_j - final\ prediction|)} \quad ...(1)$$

Where i denote initial predictors i and j denotes the other predictor. This formula captures the relative distance between the LLM's final prediction and each agent's prediction. A higher weight on predictor i indicates greater influence of predictor i on the final decision. For example, if the human agent's prediction is closer to the final prediction than the ML agent's, the human agent will receive a higher weight. This metric allowed us to objectively measure the extent to which the LLM agent relied on each source of advice.

In addition to calculating weights, we analyzed the trust scores provided by the LLM agent. These scores reflect the agent's confidence in the predictions of each individual agent and its own final prediction. By comparing trust scores with the calculated weights, we assessed whether the LLM's stated confidence aligned with its actual reliance on each agent's input. This comparison is critical for understanding discrepancies between perceived trust and decision-making behavior, particularly in hybrid human-AI systems.

We utilized a dataset from an existing study (Ibrahim et al. 2021), which originally examined human reliance on AI advice. For our experiments, we adapted this dataset to include predictions from both human and ML agents. The LLM agent was implemented using OpenAI's ChatGPT-4, a state-of-the-art model known for its reasoning and predictive capabilities. The prompts used to instruct the LLM agent were carefully designed to ensure consistency and minimize ambiguity. For example, the agent was explicitly instructed to consider the last round's performance and the true value when making its final prediction. Detailed prompts and data specifications are provided in the Appendix to facilitate reproducibility and transparency.

In summary, our methodology involved four agent combinations (H-H, H-ML, ML-ML, and Un-Un), two order configurations for each combination (e.g., H-ML and ML-H), and 200 rounds per condition to ensure robust statistical analysis. We calculated weights on advice to quantify the influence of each agent's prediction and analyzed trust scores to assess confidence and alignment with decision weights. This comprehensive approach allowed us to rigorously examine how LLM agents evaluate human and machine-generated advice, providing insights into the dynamics of human-AI collaboration.

## 4. Results

In this section, we provide the result of our econometric estimate. We create a panel data for each agent at each round level and obtained weight that is elicited by the AI agent. For each scenario, we estimate the following econometric model:

$$weight_i \sim \beta \times agent_i + lag\_error_i + \epsilon_i \ldots (2)$$

Here *weight* is the weight provided to the agent *i* and lag_error is the error of agent *i* in the previous round by agent i. In addition to *weight,* we also estimate how does the AI agent trust the agent's prediciton and how much it trusts its own prediction.

### 4.1. Both predictors have similar higher error

The Table 1 presents regression results examining how different factors influence weight, trust in an agent, and self-trust, with an unknown identity baseline (omitted reference category). The coefficients indicate deviations from this baseline. For weight in column (1), the human source (source = h) has a significant negative effect (−1.524, p<0.01), while the machine learning source (source = ml) shows no significant impact. Trust in the agent (2) is not significantly affected by either source, but self-trust (3) increases slightly at 10% level, when the source is machine learning (0.706, p<0.1).

**Table 1: Effect of identity on the weight and trust scores**

|  | weight | trust on agent | self-trust |
| --- | --- | --- | --- |
|  | (1) | (2) | (3) |
| *Intercept* | 61.183*** | 71.288*** | 79.822*** |
|  | (0.691) | (0.766) | (0.386) |
| *source = h* | −1.524** | −0.956 | 0.387 |
|  | (0.729) | (0.789) | (0.414) |
| *source = ml* | −0.192 | 0.162 | 0.706* |
|  | (0.727) | (0.790) | (0.416) |
| *lag_error* | −0.287*** | −0.456*** | −0.032*** |
|  | (0.007) | (0.008) | (0.003) |
| Num.Obs. | 7164 | 7164 | 7164 |
| R2 | 0.205 | 0.377 | 0.012 |
| R2 Adj. | 0.205 | 0.376 | 0.011 |
| RMSE | 19.91 | 20.59 | 10.63 |

*Note: * p<0.10, ** p<0.05, *** p<0.01, robust standard errors in the parentheses.*

This underscores that we need to be careful when using trust and how much actual weight is elicited by the AI agent. This disconnect is an important aspect to design H-AI system. Next in Table 2, using an interaction model, we find who is penalized more if they make error. As evident from the coefficients of *ml* and its interaction term with *lag_error* (2.782, p<0.05 and -0.075, p<0.01), only the human agent is penalized, and ml is only penalized in putting weights when the error is extremely high. There is no difference in trust or self-trust variables, which shows that what the AI agent reports in terms of its trust on the predictors vs what it actually does are strikingly different.

This is very important as many papers look at how much AI trust advice, but we show that the advice shown by the AI agent may be misleading as it elicit different weight that does not align with the trust scores. This can be explained by the guardrails that are available, while providing asnwers to question such as 'how much weight do you put on human prediction.'

**Table 2: Effect of error on the weight and trust scores**

|  | weight | trust on agent | self-trust |
|---|---|---|---|
|  | (1) | (2) | (3) |
| *Intercept* | 58.742*** | 71.403*** | 80.199*** |
|  | (1.020) | (1.188) | (0.498) |
| *source = h* | 0.993 | −1.127 | −0.220 |
|  | (1.168) | (1.337) | (0.560) |
| *source = ml* | 2.782** | 0.076 | 0.465 |
|  | (1.169) | (1.336) | (0.566) |
| *lag_error* | −0.226*** | −0.458*** | −0.042*** |
|  | (0.019) | (0.024) | (0.011) |
| *lag_error × source = h* | −0.063*** | 0.004 | 0.015 |
|  | (0.022) | (0.027) | (0.012) |
| *lag_error × source = ml* | −0.075*** | 0.002 | 0.006 |
|  | (0.022) | (0.027) | (0.012) |
| *Num.Obs.* | 7164 | 7164 | 7164 |
| *R2* | 0.206 | 0.377 | 0.012 |
| *R2 Adj.* | 0.206 | 0.376 | 0.011 |
| *RMSE* | 19.89 | 20.59 | 10.63 |

*Note: * $p<0.10$, ** $p<0.05$, *** $p<0.01$, robust standard errors in the parentheses.*

### 4.2. Both predictors have similar lower error

We find that our results remain consistent even when errors are smaller but statistically comparable, suggesting that the observed human aversion phenomenon is independent of agent error rates. As reported in Table 3, only the human agent faces penalization—similar to the high-error scenario—while the machine learning (ML) agent is penalized in weight assignment only under extremely high errors.

### 4.3. Both agents have different errors

Our findings demonstrate that the results remain consistent: the human agent is systematically penalized, particularly when placed in the second position. This penalty does not extend to the ML agent (see Table 4). Column 1 reveals that even when both agents produce different errors, the human agent consistently receives lower weights than the ML agent. Column 2 further shows that when high-error agents are in the first position, the human agent faces significant penalization (−1.570, $p < 0.05$)—a effect that intensifies when the high-error human agent is in the second position (−2.540, $p < 0.05$).

**Table 3: Effect of agent identity on the weight and trust scores**

|              | weight       | trust on agent | self-trust   |
|              | (1)          | (2)            | (3)          |
| ------------ | ------------ | -------------- | ------------ |
| *Intercept*  | 60.106***    | 75.691***      | 79.526***    |
|              | (0.563)      | (0.718)        | (0.378)      |
| *source = h* | −1.234**     | −2.359***      | 0.744*       |
|              | (0.586)      | (0.724)        | (0.388)      |
| *source = ml*| −0.183       | −1.444**       | 0.878**      |
|              | (0.584)      | (0.726)        | (0.390)      |
| *lag_error*  | −1.336***    | −2.185***      | −0.233***    |
|              | (0.031)      | (0.040)        | (0.021)      |
| Num.Obs.     | 7164         | 7164           | 7164         |
| R2           | 0.196        | 0.316          | 0.019        |
| R2 Adj.      | 0.196        | 0.316          | 0.018        |
| RMSE         | 15.71        | 18.62          | 9.95         |

*Note: * p<0.10, ** p<0.05, *** p<0.01, robust standard errors in the parentheses.*

While first-position bias (Shi et al. 2024) is well-documented, our results highlight that disclosing agent identity can exacerbate this disadvantage, especially when the human agent is positioned second. For managers designing human-AI augmented systems, these findings underscore the need to carefully consider both agent order and identity disclosure.

**Table 4: Effect of agents' identity on the weight**

|              | weight       | weight         | weight          |
|              | (1)          | (2)            | (3)             |
| ------------ | ------------ | -------------- | --------------- |
| *Intercept*  | 60.106***    | 61.018***      | 59.270***       |
|              | (0.563)      | (0.682)        | (0.921)         |
| *source = h* | −1.234**     | −1.570**       | −2.540**        |
|              | (0.586)      | (0.757)        | (1.003)         |
| *source = ml*| −0.183       | −0.073         | 0.046           |
|              | (0.584)      | (0.757)        | (1.004)         |
| *lag_error*  | −1.336***    | −0.471***      | −0.408***       |
|              | (0.031)      | (0.010)        | (0.014)         |
| Num.Obs.     | 7164         | 7164           | 3582            |
| R2           | 0.196        | 0.300          | 0.270           |
| R2 Adj.      | 0.196        | 0.300          | 0.270           |
| RMSE         | 15.71        | 20.99          | 19.66           |
| High error   | NA           | First position | Second position |

*Note: * p<0.10, ** p<0.05, *** p<0.01, robust standard errors in the parentheses.*

## 5. Discussion and Conclusion

Our study contributes to management science in three ways. First, it extends the concept of algorithmic aversion to AI agents, demonstrating that bias in human-AI collaboration is bidirectional. Second, it identifies semantic labeling (e.g., "human expert") as a lever to mitigate bias, offering actionable insights for AI interface design. Third, it exposes the limitations of trust metrics in hybrid decision-making systems, urging organizations to audit not only AI recommendations but also how AI evaluates human input.

Practically, these findings have great implications for deploying AI in high-stakes domains. For example, in healthcare, an AI system dismissing nurse triage recommendations could delay critical interventions. In finance, algorithms undervaluing human risk assessments might amplify systemic biases. By contrast, reframing human roles (e.g., "expert analyst") could foster more balanced collaboration. Our paper aid to the literature showing the not only the identity but also the position of the human agent matters. This aids to a number of papers in the area working on human-AI collaboration such as (Dellermann et al. 2019, Choudhary et al. 2023a).

As AI transitions from tools to collaborators, understanding its capacity to evaluate human input is no longer optional—it is a prerequisite for ethical, effective human-AI partnerships. Our findings underscore that algorithms are not passive arbiters; they actively shape collaboration dynamics, often in ways that mirror human biases. By interrogating these dynamics, organizations can design systems that leverage the strengths of both humans and machines, fostering trust and equity in the age of autonomous AI.

While our study focuses on prediction tasks, future research should explore dynamic, interactive settings where humans and AI iteratively revise decisions. Additionally, the ethical dimensions of AI's evaluative bias warrant scrutiny: if AI systematically devalues certain demographics (e.g., non-expert human advisors), it could entrench inequities. Transparency in AI's decision-weighting mechanisms—beyond simplistic trust scores—will be vital for accountability.